\documentclass[useAMS,usenatbib]{mn2e}

\usepackage{graphicx}
\usepackage{amssymb}
\usepackage{amsmath}

\title[Mid-infrared colour gradients in Virgo]{Mid-infrared colour gradients and the colour-magnitude relation in Virgo early-type galaxies}
\author[M. S. Clemens, P. Panuzzo, R. Rampazzo, O. Vega, A. Bressan]{M. S. Clemens,$^{1}$\thanks{E-mail:
marcel.clemens@oapd.inaf.it (MSC); alessandro.bressan@oapd.inaf.it (AB)} P. Panuzzo$^{2}$,
R. Rampazzo$^{1}$, O. Vega$^{3}$, A. Bressan$^{1,3,4}$\footnotemark[1]\\
$^{1}$INAF-Osservatorio Astronomico di Padova, Vicolo dell'Osservatorio, 5, 35122 Padova, Italy\\
$^{2}$CEA, Laboratoire AIM, Irfu/SAp, Orme des Merisiers, 
F-91191 Gif-sur-Yvette, France\\
$^{3}$INAOE, Luis Enrique Erro 1, 72840 Tonantzintla, Puebla, Mexico\\
$^{4}$SISSA-ISAS, International School for Advanced Studies, ia Beirut 4, 34014 Trieste, Italy\\
}

\begin{document}

\date{Accepted 1988 December 15. Received 1988 December 14; in original form 1988 October 11}

\pagerange{\pageref{firstpage}--\pageref{lastpage}} \pubyear{2002}

\maketitle

\label{firstpage}

\begin{abstract}

We make use of Spitzer imaging between 4 and $16\;\rm \mu m$ and near-infrared data at 
$2.2\;\rm \mu m$ to investigate the nature and distribution of the mid-infrared
emission in a sample of early-type galaxies in the Virgo cluster. These data allow us
to conclude, with some confidence, that the emission at $16\;\rm \mu m$ in passive ETGs 
is stellar in origin, consistent with previous work concluding that the excess 
mid-infrared emission comes from the dusty envelopes around evolved AGB stars. There is 
little evidence for the mid-infrared emission of an unresolved central component, as might 
arise in the presence of a dusty torus associated with a low-luminosity AGN.

We nonetheless find that the $16\;\rm \mu m$ emission is more centrally peaked than the 
near-infrared emission, implying a radial stellar population gradient. By comparing with
independent evidence from studies at optical wavelengths, we conclude that a metallicity 
that falls with increasing radius is the principal driver of the observed gradient. 

We also plot the mid-infrared colour-magnitude diagram and combine with similar work
on the Coma cluster to define the colour-magnitude relation for absolute K-band 
magnitudes from -26 to -19. Because a correlation between mass and age would produce a relation
with a gradient in the opposite sense to that observed, we conclude that the relation
reflects the fact that passive ETGs of lower mass also have a lower average metallicity. 
The colour-magnitude relation is thus driven by metallicity effects.

In contrast to what is found in Coma, we do not find any objects with anomalously bright
$16\;\rm \mu m$ emission relative to the colour-magnitude relation. Although there is 
little overlap in the mass ranges probed in the two clusters, this may suggest that 
observable ``rejuvenation'' episodes are limited to intermediate mass objects.

\end{abstract}


\section{Introduction}
\label{sec:intro}

The mid-infrared emission of even the most passive early-type galaxies (ETGs)
shows an excess of emission longward of $\sim 8\;\rm \mu m$ over that which is 
expected from purely photospheric emission. This excess emission was detected 
by Impey et al. (1986) using ground-based observations and was subsequently
studied using ISO (e.g. Bressan et al. 2001). The excess has now been observed
in the central regions of bright ETGs in the Virgo cluster (Bressan et 
al. 2006) using Spitzer-IRS. Although some ETGs show a mid-infrared excess that
is evidently caused by warm dust in a star forming inter-stellar medium 
(Panuzzo et al. 2007) the excess in passive objects can be explained as the
integrated emission from the hot dust in the envelopes of evolved AGB stars.
However, hot dust emission could also arise from the dusty torus around a 
central active galactic nucleus (AGN) of low luminosity. 

Although there is evidence that the emission at $16\;\rm \mu m$ is extended in 
Coma cluster galaxies (Clemens et al. 2009a) the weight of evidence for this is far 
from conclusive. In the present article we discuss Spitzer-IRS peakup imaging 
at $16\;\rm \mu m$ of ETGs in the Virgo cluster that permit the investigation 
of the spatial distribution of the \emph{excess} mid-infrared emission. In 
order to isolate the effects of the mid-infrared excess caused by AGB stars 
we also use IRAC images at 4.5 and $8.0\;\rm \mu m$ because these are relatively 
unaffected by the excess and sample the purely photospheric component at the 
longest possible wavelength. We will argue that the excess is stellar in origin,
as has been previously suggested (Bressan et al. 1998, Athey et al. 2002, 
Xilouris et al. 2003).

The stellar origin of the excess mid-infrared emission makes it a particularly 
useful age/metallicity diagnostic because the contribution of the mid-infrared 
excess to the integrated spectrum varies with age and metallicity in a different 
way to optical diagnostics (Bressan, Granato \& Silva 1998). As a stellar 
population gets younger and/or the metallicity {\sl increases} the mid-infrared 
excess increases. However, the optical H$\beta$ index becomes larger (optical colours bluer) 
as the age decreases and/or the metallicity {\sl decreases}. The opposite behavior 
of optical indices (colours, narrow band indices and spectral shape) and mid-IR 
excess with respect to age and metallicity variations, means that the mid-infrared 
spectral region, in combination with optical data, can disentangle age-metallicity 
effects in early-type galaxies. Mid-infrared observations, then, can provide 
{\sl without the need of accurate modeling} the explanation of the origin of the 
colour-magnitude relation in clusters of galaxies. This can be understood by considering
the two following extremes.

\begin{itemize}
\item If galaxies are coeval and the bluer colours toward the less luminous sources are 
due to  lower average metallicity (e.g. Arimoto \& Yoshii, 1986) we expect the 
mid-infrared excess to become relatively less pronounced at decreasing luminosity.

\item If the galaxies have the same average metallicity and the bluer colours toward 
less luminous sources are due to a younger age (the alternative explanation allowed 
in principle by optical diagnostics) we expect the mid-infrared excess to become 
relatively more pronounced at decreasing luminosity.
\end{itemize}
  
Although most studies tend to find that both age and metallicity vary as a function of
galaxy luminosity or velocity dispersion (e.g. Nelan et al. 2005, Thomas
et al. 2005, Clemens et al. 2006, 2009b) it is not clear which is the main driver of the 
colour-magnitude relation in the cluster environment. Here we attempt to address this 
question using a multi-wavelength appraoch. 

In Clemens et al. (2009a) we used Spitzer-IRS peakup imaging at $16\;\rm \mu m$ to
populate the luminous part of the mid-infrared colour magnitude diagram for the Coma 
cluster. In Coma, galaxies with K-band magnitudes only as faint as -22 could be studied.
In the present paper we repeat the analysis for fainter objects in the Virgo cluster
in order to extend the colour-magnitude relation to fainter magnitudes.

\section{Data analysis}
\label{sec:data}

The aperture photometry for both the IRAC and IRS-peakup images is based on post basic 
calibration data (PBCD). 

The $16\;\rm \mu m$, blue peakup images are background dominated and the background often 
shows fluctuations on spatial scales similar to the source sizes. Background subtraction
techniques that attempt to model the background as the smoothly varying component of an image
thus either include extended emission from the galaxy as background or do not correct for
the background fluctuations. This tends to cause an over-subtraction of the background in 
the vicinity of the galaxy. We therefore define the background level, at all
wavelengths, to be the median value in an elliptical annulus, of width 1.5 pixels, centred 
on the galaxy, with semi-major axis and position angle as given by the mean optical diameter 
catalogued in ``Hyperleda''\footnote{http://leda.univ-lyon1.fr}. Thus we sample the background 
at twice the radius of the galaxy. This is actually the background estimation technique 
recommended for IRAC images of extended sources\footnote{http://ssc.spitzer.caltech.edu/irac/iracinstrumenthandbook/33/}.    

Confusing sources, normally in the background, can be problematic, especially at 
$16\;\rm \mu m$. For meaningful aperture aperture photometry they need to be removed. 
In some cases the contaminating sources were too close or bright to attempt removal 
(e.g. vcc~0951). In some cases the $16\;\rm \mu m$ detections themselves were not at 
the catalogued position and so were taken as ``suspect'' background sources and 
discarded (e.g. NGC~4366). For the remaining objects that showed nearby contaminating 
sources we applied an algorithm to remove them which analyzed the pixel distribution
in a series of annuli with axis ratio and position angle equal to that of the galaxies.
Pixels that deviated by more than a fixed number of standard deviations from the median
value were replaced by the median value. The threshold for source removal was varied
between two and four standard deviations and was sometimes applied iteratively to achieve 
the best results. In all cases the results were carefully checked by eye.   

Integrated fluxes were taken as the total flux within the background annulus. We stress that 
these fluxes are measured only for the purposes of determining the mean colour of the 
galaxies. Given the way in which the background is determined, they are not a good measure of 
the true integrated fluxes of the objects. The colours of the objects, however, should be
robustly determined. When we come to plot the colour-magnitude diagram we will use the 
total K-band magnitude of the galaxies as given in the 2MASS catalogue, rather than our 
measurement from the 2MASS K-band image.   

Radial colour profiles were determined using a series of elliptical annuli with axis ratios
and position angles as given in ``Hyperleda''. Before fluxes were measured the two images
were first convolved to identical resolutions. This was achieved by convolving each image
with the measured PSF of the other. For example, when considering the K-[16] colour the
2MASS K-band image was convolved with the Spitzer blue peakup PSF and the $16\;\rm \mu m$
image was convolved with the 2MASS K-band PSF. Instrumental resolutions (FWHM) are 2\farcs6, 
1\farcs7, 1\farcs9 and 3\farcs6 for 2MASS, IRAC 4.5 and $8\;\rm \mu m$ and $16\;\rm \mu m$
respectively.    

The error on the derived fluxes was taken as the quadrature sum of the rms variation within the
aperture or annulus and the calibration error, which was taken as 5\% for the blue peakup data
and 10\% for the IRAC data (see the respective instrument handbooks).

All aperture photometry was performed using a custom pipeline written in {\sc idl} and use 
was made of the {\it IDL Astronomy User's Library} (Landsman 1995).

\section{Results}

\begin{figure*}
\centerline{
\includegraphics[scale=0.65]{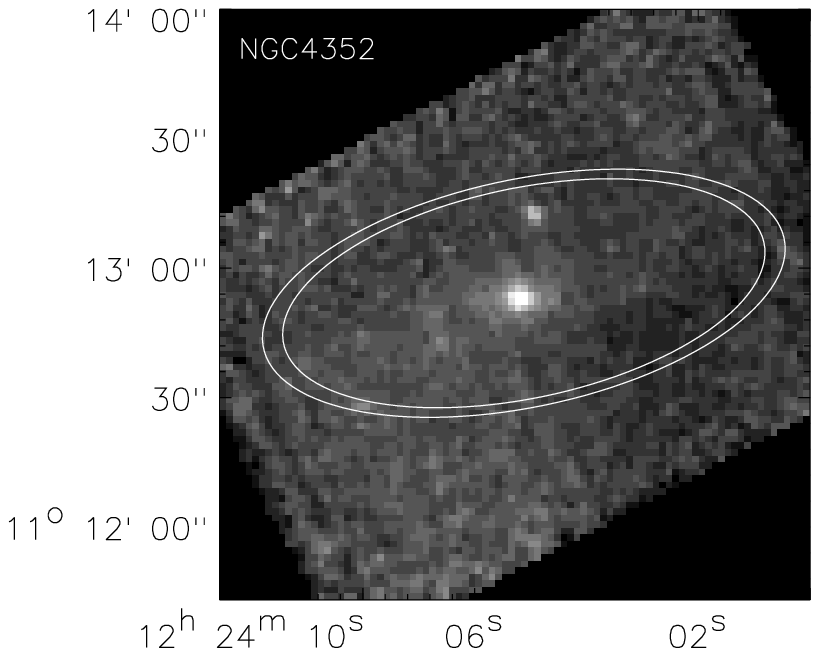}
\hskip-3mm
\includegraphics[scale=0.65]{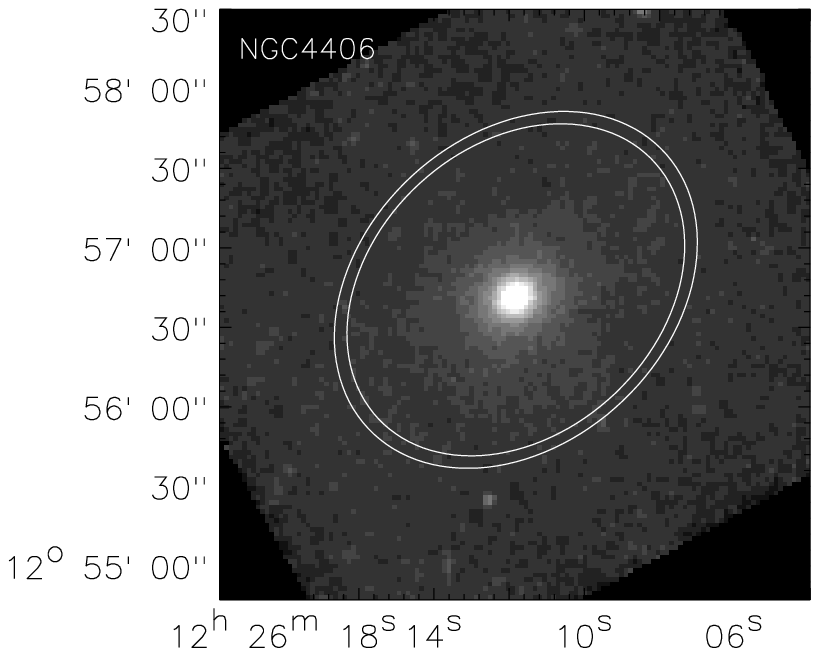}
}
\vskip-2mm
\centerline{
\includegraphics[scale=0.65]{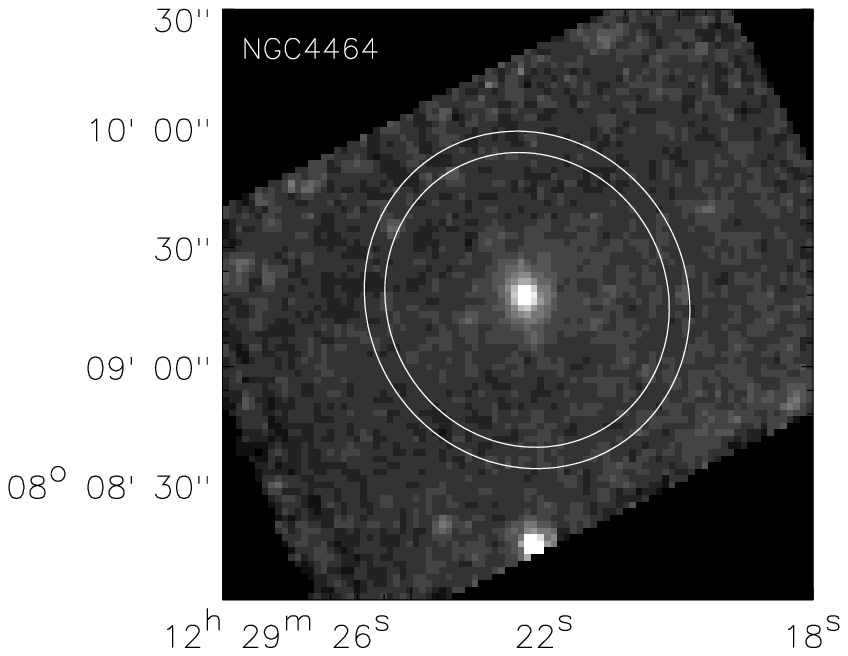}
\hskip-3mm
\includegraphics[scale=0.65]{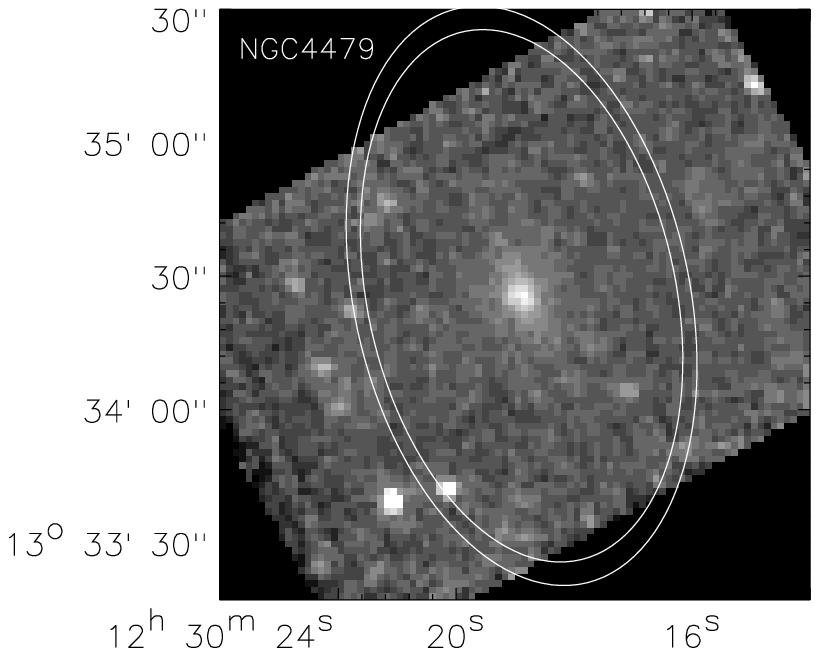}
}
\caption{Example $16\;\rm \mu m$ images. The annuli used for the determination of the 
background are shown. Background sources that can be seen in some images were removed before
the photometric measurements.}
\label{fig:images}
\end{figure*}

\begin{figure*}
\centerline{
\includegraphics[scale=0.65]{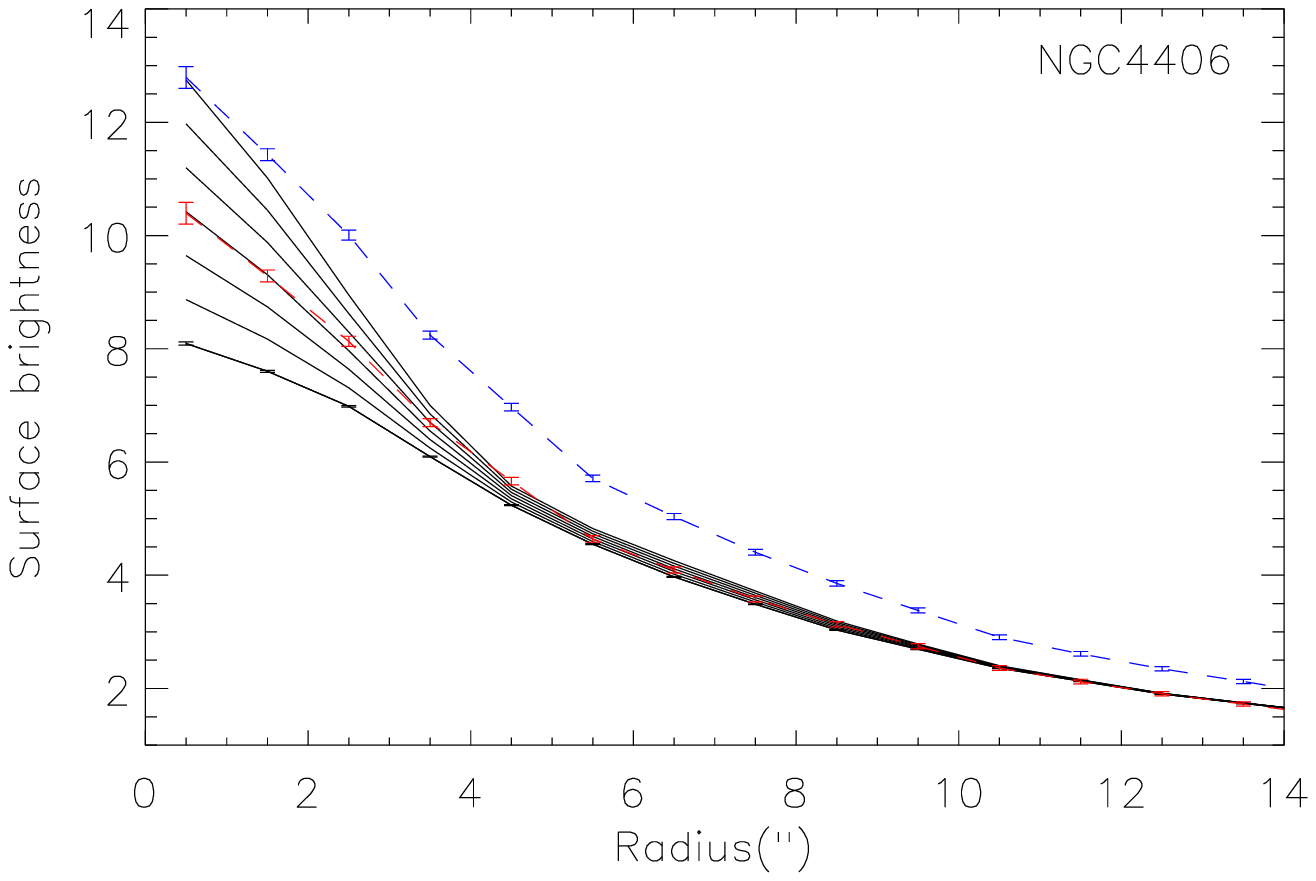}
\hskip-10mm
\includegraphics[scale=0.65]{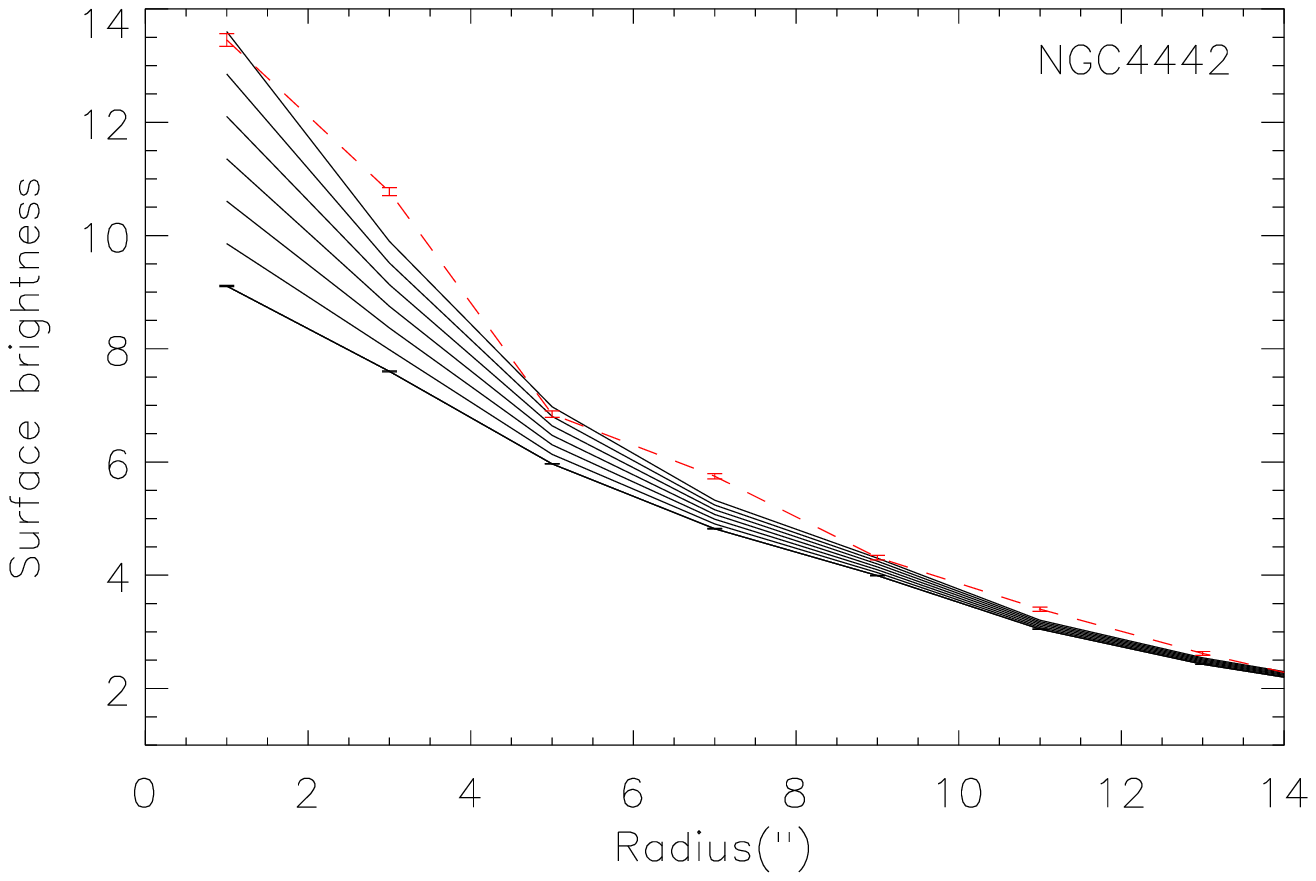}
}
\vskip-5mm
\centerline{
\includegraphics[scale=0.65]{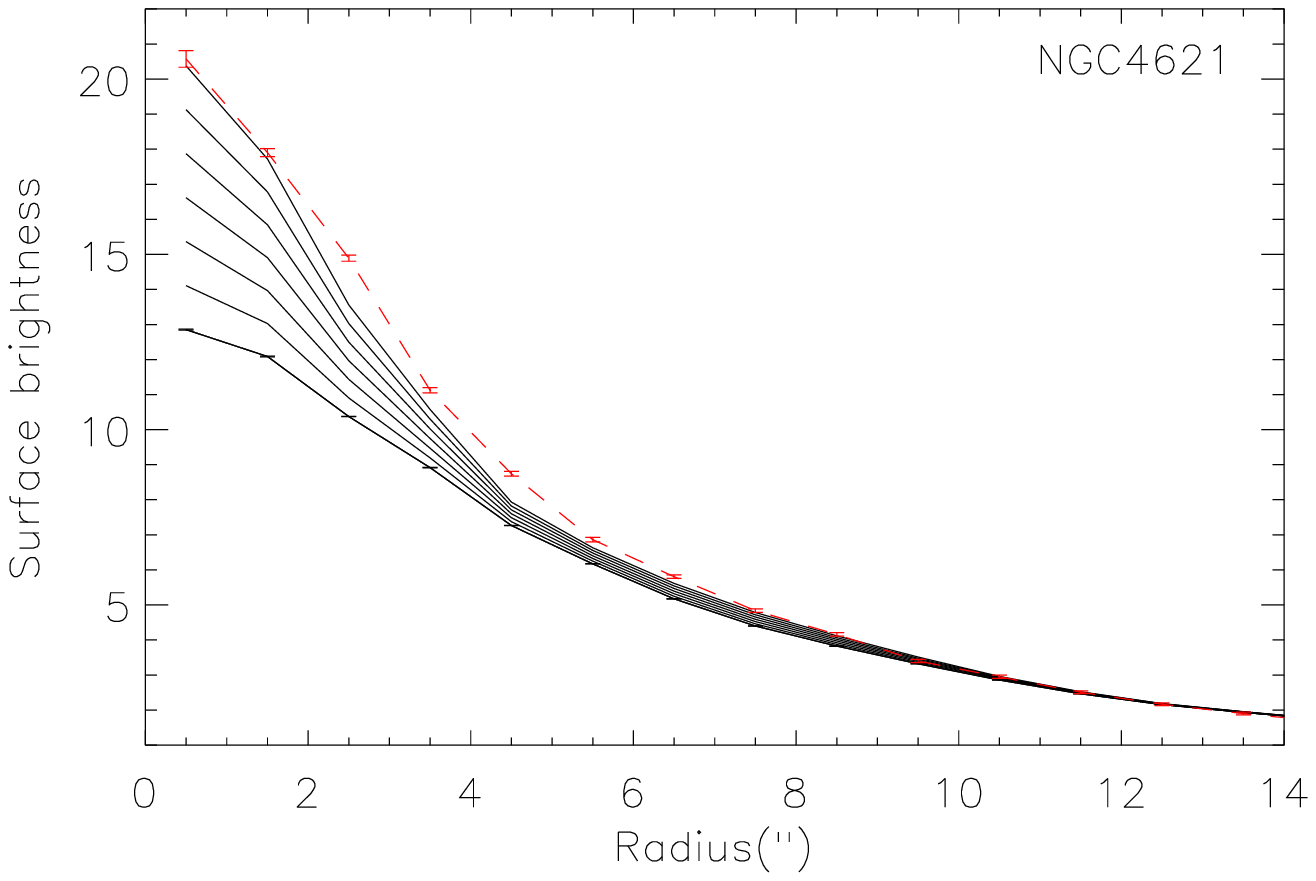}
\hskip-10mm
\includegraphics[scale=0.65]{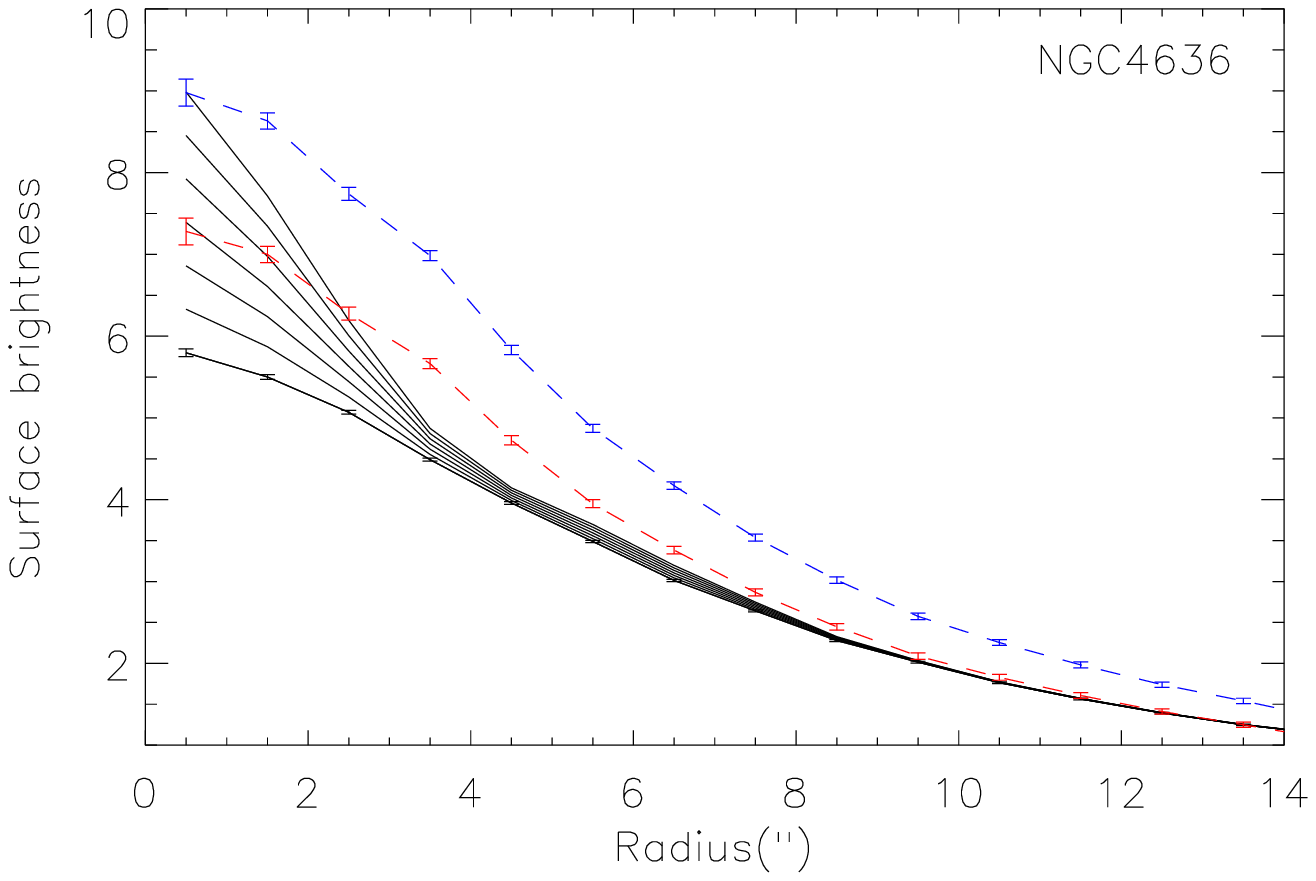}
}
\caption{Comparison of model $16\;\rm \mu m$ surface brightness profiles (solid black lines) 
with the observed profiles for the four largest galaxies in the sample (dashed lines). For the
red line the models and observed profile have been normalized at a radius of $15^{\prime\prime}$, 
well beyond any significant psf effects. The blue line, instead, is normalized to a central
point source contribution of 60\%, approximately the contribution of dusty AGB star envelopes
as measured in Spitzer-IRS spectra. The model profiles assume a $16\;\rm \mu m$ brightness 
similar to that in the K-band with an additional point source at the centre. The contribution 
of the point source rises from zero (bottom line) in increments of 10\%. Surface brightness 
units are arbitrary.}
\label{fig:point_source}
\end{figure*}

\begin{figure*}
\centerline{
\includegraphics[scale=0.9, angle=0]{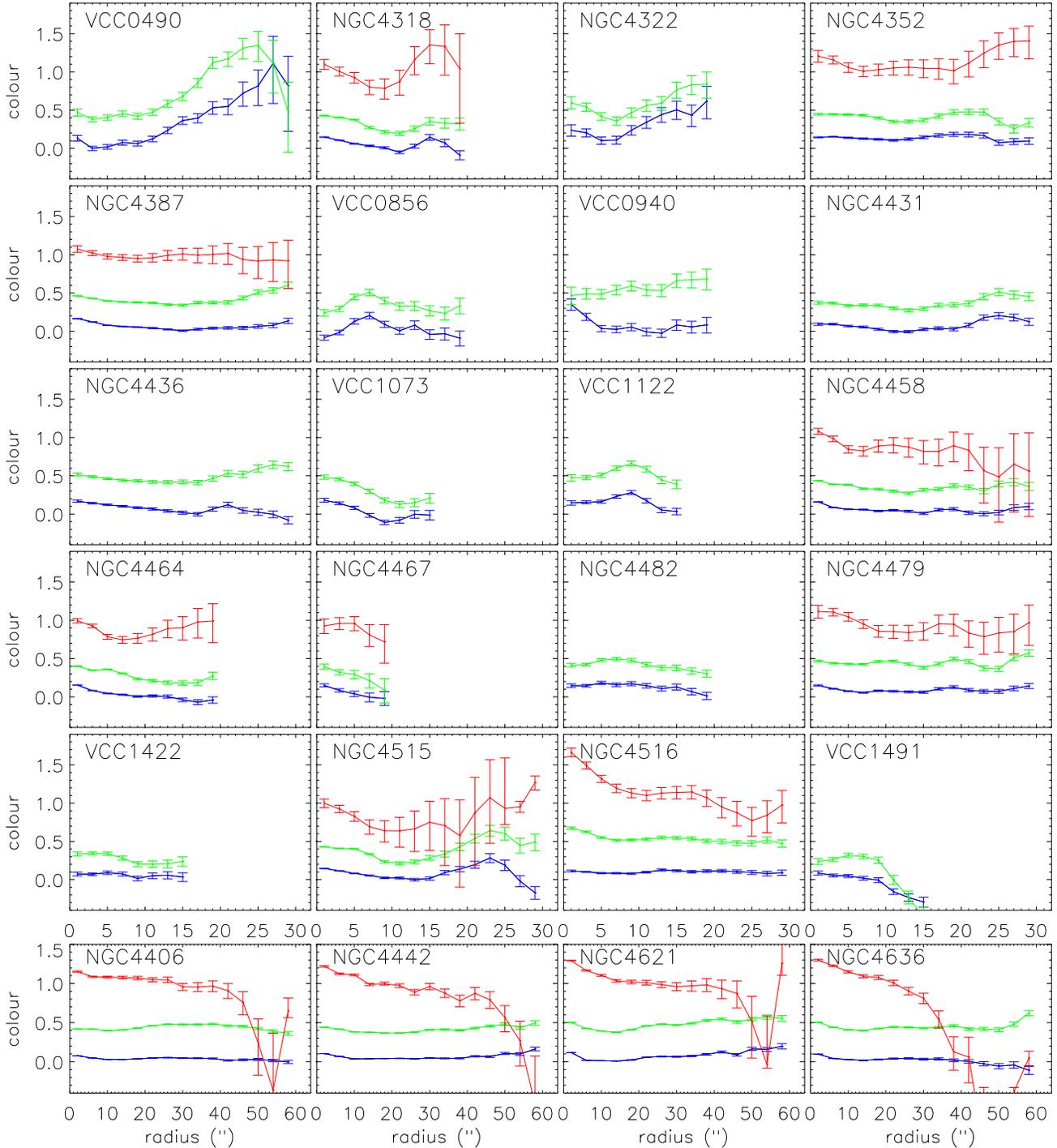}
}  
 \caption{Mid-infrared radial colour profiles. Red: $\rm K_{s}-[16]$, green:$\rm K_{s}-[8]$,
blue: $\rm K_{s}-[4.5]$.}
\label{fig:k-16}
\end{figure*}

\begin{table*}
\centering
\begin{tabular}{c c c c c c c c}
\hline\hline
Source ID & Type & Ks & $\rm K_{s}-[4.5]$ & $\rm K_{s}-[8]$ & $\rm K_{s}-[16]$ \\
          &      &    & & & \\
\hline
vcc0490 & dS0 & -20.1 &$0.15^{+0.104}_{-0.115}$ &$0.51^{+0.105}_{-0.115}$ & $-$\\
ngc4318 & E   & -21.2 &$0.06^{+0.104}_{-0.114}$ &$0.32^{+0.104}_{-0.115}$ & $1.01^{+0.079}_{-0.086}$\\
ngc4322 & dE  & -19.4 &$0.24^{+0.106}_{-0.117}$ &$0.50^{+0.106}_{-0.118}$ & $-$\\
ngc4352 & S0  & -21.4 &$0.14^{+0.104}_{-0.115}$ &$0.41^{+0.104}_{-0.115}$ & $1.18^{+0.068}_{-0.073}$\\
ngc4366 & dE  & -19.8 &$-$ &$-$ & $-$\\
vcc0786 & dE  & -19.2 &$0.06^{+0.111}_{-0.122}$ &$0.16^{+0.113}_{-0.125}$ & $-$\\
ngc4387 & E   & -22.1 &$0.07^{+0.104}_{-0.114}$ &$0.41^{+0.104}_{-0.114}$ & $0.94^{+0.065}_{-0.069}$\\
vcc0856 & dE  & -20.1 &$0.07^{+0.105}_{-0.116}$ &$0.37^{+0.105}_{-0.116}$ & $-$\\
vcc0940 & dE  & -19.4 &$0.05^{+0.105}_{-0.116}$ &$0.54^{+0.107}_{-0.118}$ & $-$\\
vcc0951 & dE  & -19.8 &$-$ &$-$ & $-$\\
ngc4431 & dS0 & -20.9 &$0.06^{+0.104}_{-0.115}$ &$0.35^{+0.104}_{-0.115}$ & $0.42^{+0.121}_{-0.137}$\\
ngc4436 & dE  & -20.5 &$0.07^{+0.104}_{-0.115}$ &$0.48^{+0.104}_{-0.115}$ & $-$\\
vcc1073 & dE  & -20.4 &$0.01^{+0.104}_{-0.115}$ &$0.28^{+0.104}_{-0.115}$ & $-0.13^{+0.217}_{-0.271}$\\
vcc1087 & dE  & -20.3 &$0.14^{+0.143}_{-0.164}$ &$0.38^{+0.255}_{-0.333}$ & $-$\\
vcc1122 & dE  & -19.4 &$0.14^{+0.104}_{-0.115}$ &$0.51^{+0.104}_{-0.115}$ & $-$\\
ngc4458 & E   & -21.8 &$0.06^{+0.104}_{-0.114}$ &$0.35^{+0.104}_{-0.115}$ & $1.03^{+0.070}_{-0.075}$\\
ngc4464 & E   & -21.7 &$0.04^{+0.104}_{-0.114}$ &$0.30^{+0.104}_{-0.114}$ & $0.81^{+0.063}_{-0.067}$\\
ngc4467 & E   & -20.2 &$0.07^{+0.105}_{-0.116}$ &$0.28^{+0.108}_{-0.120}$ & $0.86^{+0.097}_{-0.106}$\\
vcc1254 & dE  & -20.1 &$0.21^{+0.111}_{-0.122}$ &$0.50^{+0.111}_{-0.122}$ & $-$\\
ngc4482 & dE  & -20.7 &$0.13^{+0.104}_{-0.115}$ &$0.42^{+0.104}_{-0.115}$ & $0.76^{+0.097}_{-0.106}$\\
ngc4479 & S0  & -21.5 &$0.09^{+0.104}_{-0.114}$ &$0.45^{+0.104}_{-0.115}$ & $0.92^{+0.070}_{-0.075}$\\
vcc1308 & dE  & -18.5 &$0.16^{+0.108}_{-0.119}$ &$0.33^{+0.108}_{-0.120}$ & $-$\\
vcc1348 & dE  & -18.6 &$-$ &$-$ & $-$\\
vcc1422 & E   & -20.6 &$0.06^{+0.104}_{-0.115}$ &$0.27^{+0.104}_{-0.115}$ & $-$\\
vcc1453 & dE  & -19.9 &$0.15^{+0.104}_{-0.115}$ &$0.46^{+0.104}_{-0.115}$ & $0.08^{+0.256}_{-0.336}$\\
ngc4515 & E   & -21.3 &$0.08^{+0.104}_{-0.114}$ &$0.35^{+0.104}_{-0.115}$ & $0.71^{+0.094}_{-0.103}$\\
ngc4516 & S0  & -21.4 &$0.12^{+0.104}_{-0.114}$ &$0.54^{+0.104}_{-0.115}$ & $1.24^{+0.062}_{-0.066}$\\
vcc1491 & dE  & -19.2 &$-0.01^{+0.104}_{-0.115}$ &$0.21^{+0.104}_{-0.115}$ & $-0.29^{+0.350}_{-0.520}$\\
ngc4406 & E   & -25.1 &$0.04^{+0.103}_{-0.114}$ &$0.43^{+0.103}_{-0.114}$ & $1.02^{+0.054}_{-0.057}$\\
ngc4442 & S0  & -23.9 &$0.06^{+0.103}_{-0.114}$ &$0.40^{+0.103}_{-0.114}$ & $0.96^{+0.054}_{-0.057}$\\
ngc4621 & E   & -24.5 &$0.06^{+0.103}_{-0.114}$ &$0.45^{+0.103}_{-0.114}$ & $1.05^{+0.054}_{-0.057}$\\
ngc4636 & E   & -24.7 &$0.03^{+0.104}_{-0.114}$ &$0.44^{+0.104}_{-0.114}$ & $0.93^{+0.055}_{-0.058}$\\
\hline
\end{tabular}
\caption{Mean mid-infrared clours. Missing values imply either a non-detection in one band or
that the source was confused. The errors in this table include the systematic calibration errors.
These are not included in the error bars of Fig.~\ref{fig:cm}.}
\label{tab:fluxes}
\end{table*}

\subsection{Contribution from a central point source in the mid-infrared?}
\label{sec:point}

Most ETGs in our sample show negative radial gradients in the K-[16] colour, indicating 
that the mid-infrared $16\;\rm \mu m$ emission is more centrally concentrated than the 
K-band light. The IRS-spectroscopy of Bressan et al. (2006) has already shown this
emission to be extended. However, at $16\;\rm \mu m$ there are roughly similar 
contributions from the photospheric and dusty envelope components (See figure 1 of
Clemens et al. 2010) so it remains possible that only the stellar component is extended, 
while the dusty, \emph{excess}, component is unresolved. This might be expected if these 
galaxies contained a dusty torus at their centre, associated with a low luminosity AGN. 

Using HST images, Carollo et al. (1997) interpreted the central dip in the optical surface brightness of NGC~4406 
within 0\farcs4 as evidence of a nuclear dust ring. NGC~4636 is classified as a LINER/Sy~3
and has radio jets at 1.4 and 4.8~GHz (Stanger \& Warwick, 1986). Both of these objects, then, 
show evidence of AGN activity. So is there any emission from a dusty torus at $16\;\rm \mu m$?   

Although Clemens et al. (2009a) found no evidence for central point sources at $16\;\rm \mu m$ 
for the galaxies of Coma, based on effective radii estimates, the spatial resolution was 
insufficient to investigate this possibility in detail.

In order to test this possibility we created model $16\;\rm \mu m$ images based on the K-band
images for the four large galaxies in our sample. We first convolved the K-band images to
the same resolution as the observed $16\;\rm \mu m$ images, and then added a central point
source in the form of the $16\;\rm \mu m$ psf. We then measured the radial surface brightness
profiles for point sources of various strengths and compared with the observed $16\;\rm \mu m$
profiles. The results are illustrated in Fig.~\ref{fig:point_source}.

We normalized the models in two different ways. Firstly, to the observed surface brightness 
at a radius of $15^{\prime\prime}$, so as to be far from any significant PSF effects, and 
secondly so that the observed central surface brightness was equal to that of the model with
a 60\% contribution from a point source. The choice of 60\% is based on model fits to 
Spitzer-IRS profiles of passive early-type galaxies that show that approximately this fraction
of the $16\;\rm \mu m$ emission comes from the dusty circumstellar envelopes of AGB stars 
(Bressan et al. 2006). If the colour gradient is due to a population gradient involving AGB 
stars then this would be a sensible upper limit to the extent to which the colour of the 
nucleus might differ from that at larger radii.

Fig.~\ref{fig:point_source} shows that the observed profiles tend to have shallower gradients 
than one would expect if a central point source were the cause of the observed colour 
gradients. Nonetheless, the limited signal-to-noise ratio and spatial resolution of the
$16\;\rm \mu m$ images means that a central point source is permitted at some level. The most
interesting case is NGC~4636 because this contains a radio loud AGN. The observed profile in 
Fig.~\ref{fig:point_source} is clearly much shallower than would be expected if there were
a significant contribution from a central point source. At least for this galaxy, there is 
no evidence of a significant contribution to the mid-infrared emission from warm dust in a 
central torus. The observed colour gradient is therefore due to a stellar population 
gradient. If this is so for an ETG with a known AGN, there seems little evidence that the
colour gradients in the other galaxies are due to anything else. A similar conclusion was
actually reached for the mid-infrared emission of M~87, where the only excess over purely 
stellar emission is synchrotron emission from the radio loud nucleus (Buson et al. 2009, Baes et al. 2010).

\subsection{Radial colour gradients}
\label{sec:gradients}

Fig.~\ref{fig:k-16} shows the radial mid-infrared colour profiles, K-[4.5], K-[8] and
K-[16]. The signal-to-noise ratio of the  $16\;\rm \mu m$ images limits the number of
galaxies for which the K-[16] colour profile can be plotted, and in general, profiles
are truncated before noise becomes dominant. In addition, any objects that had nearby 
confusing sources have been excluded from the analysis. The 4 brightest galaxies are 
separated in the figure and have profiles that are well measured within a radius of 
$\sim 1^{\prime}$. We place more emphasis on these objects in the following discussion 
as the profiles are the most reliable. 

For the 4 brightest objects, and for the majority of the fainter galaxies, at least at 
radii less than $10^{\prime\prime}$, the K-[16] colour shows a negative radial gradient 
that is steeper than that of either the K-[4.5] or K-[8] gradients. We also see that 
(with the exception of NGC~4516) the K-[4.5] and K-[8] profiles are very nearly parallel. 
In several cases, including the brightest galaxies, these profiles are almost flat.   

With reference to the model fits to mid-infrared Spitzer-IRS spectra (Bressan et al. 2006)
we expect both the 4.5 and $8\;\rm \mu m$ micron fluxes to be dominated by emission
from stellar photospheres. However, at $16\;\rm \mu m$ this changes, with approximately
$60\%$ of the emission coming from the hot dusty envelopes around evolved AGB stars.
This offers a very natural explanation for the observed difference between the 
mid-infrared colour profiles, suggesting that the radial distribution of dusty AGB stars
is more centrally concentrated than that of the general stellar population.

\subsection{Colour--magnitude diagram}

Table~\ref{tab:fluxes} gives the mid-infrared colours of our sample galaxies determined
within the optical radius, as described in section~\ref{sec:data}.

\begin{figure}
\centerline{
\includegraphics[scale=0.36, angle=90]{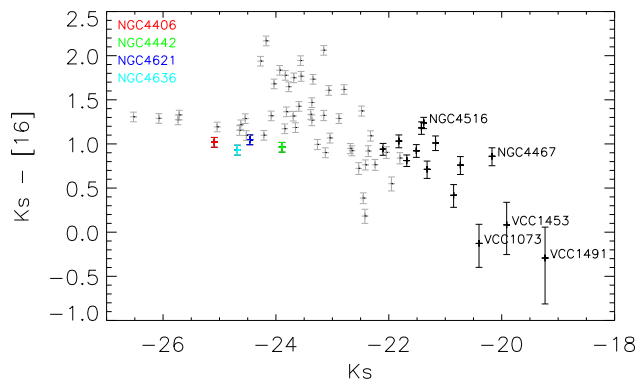}
}  
 \caption{MIR colour-magnitude diagram. The colours are defined as those within 80\%
of the optical radius given in ``Hyperleda''. K-band fluxes were taken from the
2MASS All Sky Survey.}
\label{fig:cm}
\end{figure}

In Fig.~\ref{fig:cm} we show the mid-infrared colour-magnitude diagram for those Virgo
galaxies for which the K-[16] colour could be reliably determined. In light grey we 
reproduce the values obtained for the Coma cluster by Clemens et al. (2009a). The galaxies
observed in the Virgo cluster allow this diagram to be extended to fainter magnitudes than
was possible for the Coma cluster alone, with objects down to an absolute K-band magnitude
of -19.23.  

The colour magnitude relation of the Coma cluster showed a trend for bluer K-[16] colours 
towards fainter magnitudes. The addition of the fainter Virgo galaxies shows that this trend
continues to fainter magnitudes. There appears to be a slight offset between the Coma and 
Virgo data. Although this could be caused by an error in the relative distance of the two 
clusters (taken to be 97 and 17~Mpc respectively) it may also be due to the fact that the 
colours were determined within larger apertures for the Coma cluster. The sense of the offset
(Virgo objects appearing with slightly higher than expected K-[16]) is consistent with the
colour gradients described in section~\ref{sec:gradients}. 

We also note that the 4 bright Virgo galaxies lie on the low boundary of values of K-[16] found
in the Coma cluster. There is no obvious data analysis effect that could have caused such a bias.

\section{Discussion}

Both age and metallicity can effect the strength of the 
integrated $16\;\rm \mu m$ emission from the dusty envelopes of the AGB population.
Younger stellar populations, for example, should have a larger
contribution from dusty AGB stellar envelopes. However, in order to separate the effects
of age and metallicity we will compare our results with those of studies at optical
wavelengths. As discussed in section~\ref{sec:intro}, the different responses of observables
in the optical and infrared, to changes in metallicity and age, can be used to resolve
the degeneracy between these two parameters.

\subsection{Radial colour gradients}

As discussed in section~\ref{sec:point} it seems likely that the radial gradients in the 
K-[16] colour are due to a population gradient of AGB stars relative to the general 
stellar population. The redder K-[16] colours toward the centre of the galaxies could 
be due to younger ages, higher metallicities or both.

Clemens et al. (2009b) have used optical line strength indices of a sample 
of 14,000 ETGs from the Sloan Digital Sky Survey (SDSS) to determine age and metallicity 
as a function of mass, environment and also galactic radius. They find negative radial 
metallicity gradients for all masses. Although they also find positive radial age gradients
for massive galaxies this trend is less apparent at intermediate masses and may even invert 
for low mass galaxies. The 4 bright galaxies in our Virgo cluster sample have central velocity
dispersions in the range 170-$250\;\rm km\,s^{-1}$. In this case little or no age gradient 
is expected, at least on a statistical basis. Although it must be borne in mind that the 
results based on the SDSS are statistical in nature, and not necessarily applicable to any 
individual object, these results would favour metallicity effects over age effects as the 
cause of the mid-infrared colour gradients.

Negative metallicity gradients and negligible age gradients have, in fact, been found by
other studies of ETGs both in the field (Annibali et al, 2007) and cluster (Rawle et al. 2008, 
Mehlert et al. 2003, Peletier et al. 1999) using optical line strength indices.

\subsection{Colour-magnitude relation}

Since the work of Bower, Lucey \& Ellis (1992) the colour-magnitude 
relation (its slope and narrowness) has been open to two interpretations. 
One is that ETGs are all very old and the slope of the 
colour-magnitude relation reflects a trend for lower mass galaxies to be less 
metal-rich. The narrowness results from a small dispersion in the formation epoch
with respect to the age of the system. In the second interpretation, if the 
luminosity weighted ``average age'' is allowed to vary, the only way to maintain 
a small dispersion is to invoke a synchronization mechanism, in such a way that 
lower mass ETGs appear younger.

For many years, the metallicity trend had been favoured partly because of its simplicity
and partly because of lack of evidence for a relation between mass and age 
(e.g Kodama et al. 1998). However, recently, numerous
studies have found evidence for a so-called ``downsizing'' effect (e.g. Cimatti, Daddi 
\& Rezini, 2006), which is nothing 
less than such a relation, in the sense that star formation is more prolonged
in galaxies of lower mass. The evidence of the presence of a mechanism 
that links the mean age of an object to its mass restores some of the original doubt
concerning the cause of the narrow colour-magnitude relation.

Studies at optical wavelengths, based on line strength indices, find that less massive 
galaxies tend to be younger and less metal rich (Bressan et al. 1996, Kuntschner et al. 
2001, Thomas et al. 2005, Nelan et al. 2005, Clemens et al. 2006, Smith et al. 2007), 
both in the field and in the cluster environment. If the age trend holds also for the galaxies 
in our sample (both in Virgo and in Coma) then for a fixed metallicity, we would expect 
lower mass galaxies to have more $16\;\rm \mu m$ emission per K-band flux relative to the 
more massive objects, because of the greater importance of the dusty AGB phase. That is, 
low luminosity objects should have higher values of K-[16] if age were the dominant effect. 
The observed colour-magnitude relation, however, shows the opposite trend, with lower mass 
systems having lower K-[16] colours. Metallicity, then, drives the mid-infrared 
colour-magnitude relation such that stellar populations with lower metallicity emit less 
$16\;\rm \mu m$ flux per unit stellar mass than their more metal-rich and more massive 
counterparts. In addition, population synthesis modelling of the K-[16] vs V-K colour plane 
shows that the optical colour-magnitude relation is also driven by a dominant metallicity
effect (see Fig.6 of Clemens et al. 2009a). Thus, Spitzer shows that the cluster 
colour-magnitude relation is a sequence of metallicity rather than age.   

Therefore, although a relation between mass and age could exist for our sample galaxies
this effect is dominated by a mass-metallicity trend that defines the colour-magnitude 
relation. Thus, ``downsizing'', if present, is of secondary importance in passive ETGs.

The colour-magnitude diagram of the Coma cluster shows several objects that have high
$16\;\rm \mu m$ fluxes relative to the colour-magnitude relation. None of the ETGs in
the present Virgo sample show this. Although the statistics are rather poorer for the Virgo 
sample and in Virgo we do not sample the luminosity range where these anomalous colours are 
seen in Coma, this may nonetheless indicate that the rejuvenation events that can be 
detected in the mid-infrared due to the increased $16\;\rm \mu m$ emission, are limited to 
intermediate mass objects.

\section{Conclusions}

We have observed a sample of low-luminosity ETGs in the Virgo cluster using the blue
peakup detector of Spitzer-IRS. Our aperture photometry of the spatially resolved images
can be used to draw the following conclusions.

\begin{enumerate}
\item The $16\;\rm \mu m$ emission that is in excess of that from stellar photospheres
is extended. This emission is therefore consistent with an association with the dusty
circumstellar envelopes around evolved AGB stars, rather than from a dusty torus around
a low luminosity AGN. This is a more robust conclusion than earlier evidence based on 
Spitzer-IRS spectroscopy (Bressan et al. 2006). 
\item Mid-infrared colour gradients tend to show that the $16\;\rm \mu m$ emission is
more centrally concentrated than the K-band light. Much shallower colour gradients 
involving bands dominated by photospheric emission, K-[4.5] and K-[8], imply a population
gradient in the AGB stellar population. Comparison with independent studies at optical
wavelengths implies that this is a metallicity effect. 
\item The mid-infrared colour magnitude relation, that we define over 8 magnitudes by
combining with similar data for the Coma cluster, is driven by metallicity. Less massive
objects are less metal rich.  
\item If there is a relation between galaxy mass and age in these clusters the effect
is masked by the dominant metallicity effect. Thus, the mid-infrared colour-magnitude relation,
together with optical studies, shows no evidence for so-called ``downsizing'' in the 
cluster environment. 
\item The lack of objects with K-[16] colours that place them well-above the 
colour-magnitude relation of passive ETGs in the present Virgo sample, and for the most 
massive objects in the Coma sample of Clemens et al. (2009a), may be evidence that 
observable rejuvenation episodes are limited to intermediate mass objects. 

\end{enumerate}

\section*{Acknowledgments}
MC, AB and RR acknowledge support from contract ASI/INAF I/016/07/0.

We acknowledge the usage of the HyperLeda database (http://leda.univ-lyon1.fr).

This work is based on observations made with the Spitzer Space Telescope, which
is operated by the JPL, Caltech under a contract with NASA.

We make use of data products from the Two Micron All Sky
Survey, which is a joint project of the University of Massachusetts and
the Infrared Processing and Analysis Center/California Institute of
Technology, funded by the National Aeronautics and Space Administration
and the National Science Foundation. 

This research has made use of the GOLD Mine Database.

\bsp

\label{lastpage}

\end{document}